\documentclass[fleqn,usenatbib]{mnras}

\usepackage{newtxtext,newtxmath}

\usepackage[T1]{fontenc}
\usepackage{ae,aecompl}

\usepackage{graphicx}	
\usepackage{amsmath}	

\usepackage{bm}

\renewcommand{\eqref}[1]{Eq.\,(\ref{#1})}
\newcommand{\sub}[1]{_{\rm #1}}
\newcommand{\pderiv}[2]{\frac{\partial #1}{\partial #2}}
 \newcommand{\dd}{\mathrm{d}}
 \newcommand{\brafracket}[2]{\left( \frac{#1}{#2} \right)}
 \newcommand{\bra}{\left(}
 \newcommand{\ket}{\right)}
 \newcommand{\braa}{\left[}
 \newcommand{\kett}{\right]}

\newcommand{\SigdotPE}{\dot{\Sigma}_{\rm PEW}}
\newcommand{\SigdotX}{\dot{\Sigma}_{\rm X}}
\newcommand{\SigdotEUV}{\dot{\Sigma}_{\rm EUV}}
\newcommand{\SigdotDW}{\dot{\Sigma}_{\rm MDW}}

\newcommand{\cs}{c\sub{s}}
\newcommand{\arp}{\overline{\alpha_{r\phi}}}
\newcommand{\apz}{\overline{\alpha_{\phi z}}}
\newcommand{\apzo}{\overline{\alpha_{\phi z,0}}}
\newcommand{\Mdot}{\dot{M}}
\newcommand{\Mdotacc}{\Mdot\sub{acc}}
\newcommand{\MdotPE}{\Mdot\sub{PEW}}
\newcommand{\MdotDW}{\Mdot\sub{MDW}}
\newcommand{\Mstar}{M_\star}
\newcommand{\Lstar}{L_\star}
\newcommand{\LX}{L\sub{X}}

\newcommand{\pEUV}{\Phi\sub{EUV}}
\newcommand{\Msun}{\rm M_{\sun}}

\newcommand{\Lsun}{\rm L_{\sun}}

\newcommand{\amu}{\mathrm{m_u}}
\newcommand{\kB}{\mathrm{k_B}}
 \newcommand{\au}{{\rm au}}
 \newcommand{\yr}{{\rm yr}}
 \newcommand{\Myr}{{\rm Myr}}
 \newcommand{\K}{{\rm K}}
  \newcommand{\Tmid}{T\sub{mid}}
  
 \newcommand{\cw}{C\sub{w}}
 \newcommand{\cwo}{C\sub{w,0}}
 \newcommand{\cwe}{C\sub{w,e}}
 \newcommand{\erad}{\epsilon\sub{rad}}
 \newcommand{\tnir}{t\sub{NIR}}

 \newcommand{\strtrq}{C}
 \newcommand{\str}{D}
 \newcommand{\nodw}{A}
 \newcommand{\nope}{B}
 
 \newcommand{\weak}{F}
\defcitealias{Suzuki+16}{S16}

\title[Disk Dispersal by MHD and photoevaporative winds]{
Dispersal of protoplanetary disks by the combination of magnetically driven and photoevaporative winds}

\author[M. Kunitomo et al.]{
Masanobu Kunitomo,$^{1}$\thanks{E-mail: kunitomo.masanobu@gmail.com (MK)}
Takeru K. Suzuki$^{2}$
and Shu-ichiro Inutsuka$^{3}$
\\
$^{1}$Department of Physics, School of Medicine, Kurume University, 67 Asahimachi, Kurume, Fukuoka 830-0011, Japan\\
$^{2}$School of Arts \& Sciences, The University of Tokyo, 3-8-1, Komaba, Meguro, Tokyo 153-8902, Japan\\
$^{3}$Department of Physics, Nagoya University, Furo-cho, Chikusa-ku, Nagoya, Aichi 464-8602, Japan
}

\date{Accepted 2020 January 7. Received 2019 December 6; in original form 2019 October 1}

\pubyear{2020}

\begin{document}
\label{firstpage}
\pagerange{\pageref{firstpage}--\pageref{lastpage}}
\maketitle

\begin{abstract}
We investigate the roles of magnetically driven disk wind (MDW) and thermally driven photoevaporative wind (PEW) in the long-time evolution of protoplanetary disks.
We start simulations from the early phase in which the disk mass is $0.118\,\Msun$ around a $1\,\Msun$ star and track the evolution until the disk is completely dispersed. 
We incorporate the mass loss by PEW and the mass loss and magnetic braking (wind torque) by MDW, in addition to the viscous accretion, viscous heating, and stellar irradiation.
We find that MDW and PEW respectively have different roles: 
magnetically driven wind ejects materials from an inner disk in the early phase, whereas photoevaporation has a dominant role in the late phase in the outer ($\gtrsim1$\,au) disk.
The disk lifetime, which depends on the combination of MDW, PEW, and viscous accretion, shows a large variation of $\sim1$--20\,Myr; the gas is dispersed mainly by the MDW and the PEW in the cases with a low viscosity and the lifetime is sensitive to the mass-loss rate and torque of the MDW, whereas the lifetime is insensitive to these parameters when the viscosity is high. 
Even in disks with very weak turbulence, the cooperation of MDW and PEW enables the disk dispersal within a few Myr.  
\end{abstract}





\section{Introduction}
\label{sec:intro}

Observations have revealed that the typical lifetime of protoplanetary disks is several million years (Myr) \citep[e.g.,][]{Haisch+01, Mamajek09, Yasui+10, Takagi+14}.
Their evolution and dispersal is crucial for the planet formation in protoplanetary disks: for example, 
the growth of solid particles and the migration of (proto)planets depend on the disk properties \citep[e.g.,][]{Matsuyama+03b, Ogihara+15, Kobayashi+Tanaka18}.

A current major scenario of the evolution of protoplanetary disks is that the disk material is dispersed by the combined effect of the viscous accretion and the photoevaporation (PEW hereafter) driven by the irradiation of high-energy photons from a central star \citep[e.g.,][]{Clarke+01, Alexander+06b, Gorti+09, Owen+10}.
In the early phase viscous accretion dominates the PEW and both the disk mass and accretion rate onto a star decrease with time. When the accretion rate becomes comparable to the PEW rate, a gap opens at $\sim1\,\au$ \citep[][]{Clarke+01,Liffman03}.
An inner disk is drawn by the central star quickly and the outer disk is also quickly dispersed by the PEW.

Magnetorotational instability \citep[MRI hereafter;][]{Velikhov59,Chandrasekhar61,Balbus+Hawley91} was highlighted as a source of the turbulent viscosity that induces the mass accretion. However, both recent observations \citep[e.g.,][]{Pinte+16, Flaherty+17} and theoretical studies \citep[e.g.,][]{Turner+14, Mori+16} have suggested that MRI turbulence may not be vigorous in protoplanetary disks because of the insufficient ionization.
In the MRI-inactive disks, however, the expected lifetime of protoplanetary disks is much longer ($\gtrsim 10$ Myr) than the observed value owing to the suppressed accretion, even though the PEW  simultaneously contributes to their dispersal \citep{Morishima12}.

In addition to the PEW, magnetically driven disk wind (hereafter MDW) is also one of the essential processes that controls the evolution of protoplanetary disks.
\citet{Suzuki+Inutsuka09} proposed that the MDW driven by the MRI turbulence potentially contributes to the mass loss of the gas component of protoplanetary disks.
{Follow-up studies \citep{Bai+Stone13a,Lesur+13,Fromang+13} have confirmed the MDW.}
While this mechanism is partially suppressed in magnetically inactive dead zones \citep{Suzuki+10}, magnetocentrifugal driven winds, which carry off the angular momentum of disks in addition to the mass \citep{Blandford+Payne82,Pelletier+Pudritz92}, also give a significant impact on their evolution, particularly because mass accretion is induced by the wind torque \citep{Bai+Stone13b,Simon+15,Hasegawa+17}. 
On this basis, \citet[][hereafter \citetalias{Suzuki+16}]{Suzuki+16} constructed a model that incorporates the general properties of the MDW and investigated the long-term evolution of protoplanetary disks.

In this paper we wish to investigate the disk evolution around low-mass stars\footnote{
We note that \citet{Tanaka+17} investigated the disk evolution including both winds in the context of massive star formation.
In this work we focus only on the disks around a $1\,\Msun$ star.
} including viscous accretion, the PEW and the MDW particularly focusing on the MRI-inactive disks.

This paper is organized as follows. In Sect.\,\ref{sec:method}, we describe our physical models of the PEW, the MDW and accretion, and computation method for simulating the disk evolution. In Sect.\,\ref{sec:result}, we explore how the wind model and the viscosity affect the disk evolution.
We describe the caveats of our model in Sect.\,\ref{sec:discussion}, and our results are summarized in Sect.\,\ref{sec:conclusion}.

\section{Methods} \label{sec:method}

We simulate the time evolution of protoplanetary disks including the effects of viscous accretion, PEW and MDW (Sect.\,\ref{sec:basiceq}).
We adopt wind models in the literature (Sects.\,\ref{sec:dw}, \ref{sec:pe}).
The magnetic braking by the MDW is also included (Sect.\,\ref{sec:apz}).
The settings are mostly the same as those of \citetalias{Suzuki+16} except for the PEW model (Sects.\,\ref{sec:init}, \ref{sec:setting}).

\subsection{Basic equations}\label{sec:basiceq}

We numerically solve the one-dimensional diffusion equation {\citep[e.g.,][]{Lynden-Bell+Pringle74}}:
\begin{align}
&\frac{\partial \Sigma}{\partial t} - \frac{1}{r}\frac{\partial}{\partial r}
\left[\frac{2}{r\Omega}\left\{\frac{\partial}{\partial r}(r^2 \Sigma
  \overline{\alpha_{r\phi}}c_{\rm s}^2) + r^2 \overline{\alpha_{\phi z}}
  (\rho c_{\rm s}^2)_{\rm mid} \right\}\right]\nonumber\\
&+ \SigdotDW + \SigdotPE =0\,,
  \label{eq:sgmevl}
\end{align}
with usual notations, that is, $\Sigma\equiv \int_{-\infty}^{\infty}  \rho\dd z$ is the surface density, $r$ the distance from the central star, $\Omega$ the angular velocity, $\rho$ the density, and $\cs$ the sound speed.
We use the cylindrical coordinates ($r, \phi, z$).
We neglect the disk self-gravity and the gas pressure gradient force, and assume $\Omega=\Omega\sub{K}=\sqrt{G\Mstar/r^3}$, where $G$ is the gravitational constant and $\Mstar$ the stellar mass.
The subscript ``mid'' stands for quantities at the disk midplane.
We describe the models of the mass-loss rate due to MHD ($\SigdotDW$) and photoevaporative ($\SigdotPE$) winds
later in Sects.\,\ref{sec:dw} and \ref{sec:pe}.

The terms including $\arp$ and $\apz$ represent the viscous and wind-driven accretion, respectively.
They are defined as
\begin{equation}
  \int \dd z \left(\rho v_r \delta v_{\phi} - \frac{B_rB_{\phi}}{4\pi}\right)
  \equiv \int \dd z \rho \alpha_{r\phi} c_{\rm s}^2 \equiv \Sigma
  \overline{\alpha_{r\phi}} c_{\rm s}^2
  \label{eq:rphistress}
\end{equation}
and
\begin{equation}
  \left(\rho \delta v_{\phi} v_z - \frac{B_{\phi} B_z}{4\pi}\right)_{\rm w}
  \equiv (\rho c_{\rm s}^2\alpha_{\phi z})_{\rm w} \equiv
  (\rho c_{\rm s}^2)_{\rm mid} \overline{\alpha_{\phi z}}\,,
  \label{eq:phizstress}
\end{equation}
where $\bm{B}$ represents the magnetic field, $\bm{v}$ the velocity, and $\delta v_{\phi}$ the deviation from the Keplerian velocity (see \citetalias[][]{Suzuki+16}).
The subscript ``w'' stands for quantities in wind regions.
We note that $\arp$ corresponds to the $\alpha$ viscosity introduced by \citet{Shakura+Sunyaev73}, which is related to viscosity, $\nu = (2/3) \arp \cs^{2}/\Omega$ {\citepalias[see][for the origin of the factor 2/3]{Suzuki+16}}.
We do not specify the origin of the anisotropic stress, $\arp$. In regions with sufficient ionization, MRI is the primary mechanism to give an order of $\arp\sim 10^{-2}$ \cite[e.g.,][]{Sano+04, Suzuki+10}.
On the other hand, in dead zones with insufficient ionization \citep{Gammie96}, a moderate level of $\arp\sim 10^{-4}$--$10^{-3}$ is probably sustained by hydrodynamical processes such as vertical shear instability \citep[e.g.,][]{Nelson+13}.
We describe the models of $\apz$ in Sect.\,\ref{sec:apz}.

\subsection{Disk thermal structure} \label{sec:Tprof}
For the temperature profile at the midplane, we consider viscous heating and stellar irradiation as
\begin{align}\label{eq:Tmid}
T\sub{mid}^4 &= T\sub{vis}^4 + T\sub{irr}^4
\end{align}
where
\begin{align}
&2\sigma\sub{SB}T\sub{vis}^4 = \bra \frac{3}{8}\tau\sub{R} + \frac{1}{2\tau\sub{P}} \ket F\sub{rad}\label{eq:Tvis}
\end{align}
and
\begin{align}\label{eq:Tirr}
&T\sub{irr} = 280\,\K \left( \frac{r}{\au} \right)^{-1/2} \left( \frac{ \Lstar }{ \Lsun } \right)^{1/4} \,.
\end{align}
{We refer to \citet[][]{Nakamoto+Nakagawa94} for Eqs.\,(\ref{eq:Tmid}) and (\ref{eq:Tvis}) and \citet[][]{Hayashi81} for Eq.\,(\ref{eq:Tirr}).}
We note that the ambient radiation field is often included in the literature, which sets a lower limit at around 10\,K. However, we confirmed that it has little impact on the long-term disk evolutions and we neglect the effect in this article.

The viscous heating rate, $F\sub{rad}$, is described below (see Sect.\,\ref{sec:cw}).
$\sigma\sub{SB}$ is the Stefan-Boltzmann constant.
Here we assume that the gravitational energy, which is liberated by the accretion at the midplane, is transported to the disk surface by diffusion\footnote{
We note that recent MHD simulations have revealed that this widely-used assumption may not be valid \citep{Mori+19}.
In this article we stick to this assumption for simplicity.
}.
The Rosseland-mean and Plank-mean optical depths at the midplane are given by\footnote{
We note that in \citetalias{Suzuki+16}, $\tau\sub{R}$ was defined as $\kappa\sub{R}\Sigma/2$ (see their Eq.\,24). However, the factor 1/2 was not needed \citep[see Appendix A of][]{Nakamoto+Nakagawa94}. }
\begin{align}
        &{\tau\sub{R} = \kappa\sub{R}\Sigma} \label{eq:tauR}
    \end{align}
    and
    \begin{align}
    &\tau\sub{P}=\max(2.4\tau\sub{R}, 0.5)\,. \label{eq:tauP}
\end{align}
According to \citet{Nakamoto+Nakagawa94}, the Rosseland-mean opacity, $\kappa\sub{R}$, is 
\begin{equation}\label{eq:kap}
\kappa\sub{R}=
    \begin{cases}
      4.5\,\brafracket{\Tmid}{150\,{\rm{K}}}^2\,{\rm{cm^{2}\,g^{-1}}} & {\rm for}\ T<150\,\rm{K}\\
      4.5\,{\rm{cm^{2}\,g^{-1}}} & {\rm for}\ 150\leq T\leq1500\,\rm{K}\\
      0\,{\rm{cm^{2}\,g^{-1}}} & {\rm for}\ T>1500\,\rm{K} \,.
    \end{cases}
\end{equation}
To avoid numerical problems, we use the following smoothed opacity instead of Eq.\,(\ref{eq:kap}):
\begin{align} \label{eq:kap-smth}
\kappa\sub{R} = & 2.25\,{\rm{cm^{2}\,g^{-1}}} 
\braa 1-\tanh\brafracket{T\sub{mid}-1500\,\K}{150\,\K} \kett \nonumber \\
&\times \min\braa 1, \brafracket{\Tmid}{150\,\K}^2 \kett  \,.
\end{align}
{We note that in the high-temperature ($T>1500\,\K$) range, the opacity depends on temperature in a more complicated manner \citep{Zhu+09} than that in \citet{Nakamoto+Nakagawa94}, which may cause thermal instability in the disk innermost region \citep{Bell+Lin94,Kimura+Tsuribe12}. However, since it is beyond the scope of this study to investigate the detailed properties of the thermal instability, we adopt the simpler opacity.
}

With the midplane temperature, the sound speed, the disk scale height, and the density at the midplane are given by
\begin{align}
&{c\sub{s}^2} = \frac{ \kB T\sub{mid}}{\mu \amu}\,,\\
&h = \sqrt{2}c\sub{s}/\Omega\sub{K}\,, \label{eq:h}\\
&\rho\sub{mid} = \frac{\Sigma}{\sqrt{\pi}h}\,,
\end{align}
where $\mu=2.34$ is the mean molecular weight, $\amu$ the atomic mass unit, and $\kB$ the Boltzmann constant.

\subsection{Models of magnetically driven disk winds} \label{sec:dw}

We adopt the MDW model in \citetalias{Suzuki+16}.
Here we briefly summarize the parameters of the MDW.
For full details of the model, readers are advised to refer to \citetalias{Suzuki+16}.

\subsubsection{Mass-loss rate} \label{sec:cw}

We give the the mass-loss rate of the MDW as
\begin{equation}
\SigdotDW=(\rho v\sub{z})\sub{w}=(\rho \cs)\sub{mid} \cw\,. \label{eq:DW}
\end{equation}
$\cw$ is a non-dimensional factor given by
\begin{equation}
C\sub{w}=\min \bra C\sub{w,0}, \cwe \ket\,,
\end{equation}
where $\cwo$ is a constant maximum value inferred from local shearing box MHD simulations \citep{Suzuki+Inutsuka09,Suzuki+10} and $\cwe$ the limiter from the accretion energetics of protoplanetary disks.
$\cwo$ weakly depends on the coupling between the magnetic field and the gas; $\cwo$ is smaller in magnetically inactive conditions.

Following \citetalias{Suzuki+16}, here we consider two cases for $\cwe$.
The first case (called ``strong DW'' in \citetalias{Suzuki+16}) corresponds to {the most energetic wind}, that is, all liberated gravitational energy is transferred into the energy to launch wind flows, and the viscous heating is balanced to radiative cooling, $F\sub{rad}$.
Therefore we adopt
\begin{align}
    &\cwe = \max\braa \frac{2}{r^3\Omega (\rho \cs)\sub{mid}} \pderiv{}{r}\bra r^2\Sigma\arp \cs^2 \ket + \frac{2\cs}{r\Omega}\apz , 0 \kett \,,
    \label{eq:strongcwe} \\
    &F\sub{rad} = \max \braa  -\frac{1}{r}\pderiv{}{r}\bra r^2\Sigma\Omega \arp\cs^2 \ket , 0\kett\,. \label{eq:strongfrad}
\end{align}

For the second case called ``weak DW'', we introduce a dimensionless free parameter $\erad$. 
We assume that the fraction $\erad$ of all the available energy (i.e., both gravitational energy and viscous heating) is radiated away and the rest is used to launch winds.
Therefore we obtain
\begin{align}
  &\cwe = (1-\erad)\braa
  \frac{3\sqrt{2\pi}\cs^2}{r^2\Omega^2} \arp
    + \frac{2\cs}{r\Omega} \apz
    \kett\,,
  \label{eq:weakcwe} \\
  &F\sub{rad} = \erad \braa  
    \frac{3\sqrt{2\pi}(\rho\cs^3)\sub{mid}}{ 2 } \arp
    + r\Omega(\rho\cs^2)\sub{mid} \apz
    \kett\,. \label{eq:weakfrad}
  \end{align}
In the weak MDW case, we {adopt $\erad=0.9$ as a lower limit in the wind energy}, that is, only a small amount of energy is used for winds and the rest is radiated away.
We note that in the absence of the MDW (i.e., $\erad=1$ and $\apz=0$), $F\sub{rad}$ becomes $\frac{3}{2}(\Sigma\Omega\cs^2)\arp$ that is used in standard accretion disk models.

\subsubsection{Magnetic braking}\label{sec:apz}

Following \citetalias{Suzuki+16}, we refer to the angular momentum transport (i.e., magnetic braking) by the MDW as wind torque.
Although the strength of the wind torque depends on the net vertical magnetic field \citep{Bai13}, its evolution is still uncertain \citep[see, e.g.,][]{Okuzumi+14, Takeuchi+Okuzumi14, Guilet+Ogilvie14}.
Following \citetalias{Suzuki+16}, we adopt the upper limit of the wind torque, that is, the field strength is conserved.
In this case, the relative strength of magnetic stress to gas pressure increases with decreasing $\Sigma$ and then the density-dependent $\apz$ is given by (\citetalias{Suzuki+16})
\begin{equation}
    \apz = \min \braa {\apzo} \brafracket{\Sigma}{\Sigma\sub{ini}}^{-0.66}  , 1 \kett\,, \label{eq:apz}
\end{equation}
{where $\apzo = 10^{-5}$}.

\subsection{Photoevaporation models} \label{sec:pe}

So far a number of studies have been conducted on the PEW driven by stellar irradiation \citep[e.g.,][]{Hollenbach+94,Ercolano+08,Gorti+Hollenbach09,Tanaka+13}.
We also refer to recent reviews \citep[][and references therein]{Alexander+14,Gorti+16,Ercolano+Pascucci17}.
We adopt the PEW rate, $\SigdotPE$, in the literature.
We consider the PEW by extreme ultra-violet photons (EUV; 13.6--100\,eV) and X-rays ($\geq0.1$\,keV) from a central star and assume that $\SigdotPE=\SigdotEUV + \SigdotX$.
In this paper we do not consider the external irradiation by a nearby massive star \citep[e.g.,][]{Adams+04}.
We also consider that for both EUV and X-ray PEW, the mass-loss profiles change after an inner hole is created and then the outer disk is directly irradiated by high-energy photons (so-called direct photoevaporation).

\subsubsection{Mass-loss rate}\label{sec:PEMdot}

We adopt the model of $\SigdotEUV$ for primordial disks in \citet[][see their Appendix\,A]{Alexander+Armitage07}, which is based on hydrodynamic simulations by \citet{Font+04}, and the model in \citet{Alexander+06a} for the direct PEW.
We adopt the $\SigdotX$ models for both primordial disks and the direct PEW in \citet[][see their Appendix\,B]{Owen+12}, which are based on hydrodynamic simulations by \citet{Owen+10,Owen+11b}.

The total mass-loss rates for the X-ray PEW in the both regimes are
\begin{align}
&\Mdot\sub{X,p} = 6.3\times 10^{-9}\,{\Msun/\rm{yr}}
\brafracket{\LX}{10^{30}\,\rm{erg/s}}^{1.14}
\brafracket{\Mstar}{\Msun}^{-0.068} \label{eq:Xprim}
\end{align}
and
\begin{align}
&\Mdot\sub{X,d} = 4.8\times 10^{-9}\,{\Msun/\rm{yr}}
\brafracket{\LX}{10^{30}\,\rm{erg/s}}^{1.14}
\brafracket{\Mstar}{\Msun}^{-0.148}\,, \label{eq:Xdirect}
\end{align}
where $\LX$ is the stellar X-ray luminosity, and the subscripts ``p'' and ``d'' stand for the primordial disk case and the direct PEW case, respectively.
The total mass-loss rates for the EUV PEW in the both regimes are
\begin{align}
&\Mdot\sub{EUV,p} = 1.6\times 10^{-10}\,{\Msun/{\rm{yr}}}
\brafracket{\Phi\sub{EUV}}{10^{41}{\rm{s}}^{-1}}^{1/2}
\brafracket{\Mstar}{1\Msun}^{1/2} \label{eq:EUVprim}
\end{align}
and
\begin{align}
&\Mdot\sub{EUV,d} = 1.3\times 10^{-9}\,{\Msun/{\rm{yr}}}
\brafracket{\Phi\sub{EUV}}{10^{41}{\rm{s}}^{-1}}^{1/2} \brafracket{R\sub{hole}}{3\,\au }^{1/2}\,, \label{eq:EUVdirect}
\end{align}
where $\Phi\sub{EUV}$ the EUV photon luminosity. 
We assume the aspect ratio $h/r=0.05$ in Eq.\,(\ref{eq:EUVdirect}) \citep[see][]{Alexander+06a}.
We refer to \citet{Alexander+Armitage07} and \citet{Owen+12} for the formulae of $\SigdotX$ and $\SigdotEUV$.

The $\SigdotX$ and $\SigdotEUV$ profiles for full disks have a characteristic feature:
they have a peak at $\simeq2.5\,{\au}\,(M_{\star}/\Msun)$ and $\simeq1.1\,{\au}\,(M_{\star}/\Msun)$, respectively.
The peak radii roughly correspond to the critical radius, inside which the gravitational potential is so deep that heated gas cannot flow out.
The critical radius is $R\sub{crit}\sim0.1$--$0.2 R\sub{g}$ \citep[][]{Liffman03} where $R\sub{g}$ is the gravitational radius
\begin{eqnarray}
R_{\rm{g}}&\equiv&\frac{GM_{\star}}{c_{\rm{s}}^2}
=8.87\,{\au}
\left( \frac{M_{\star}}{1\,\Msun} \right) 
\left( \frac{c_{\rm{s}}}{10\,\rm{km/s}} \right)^{-2}\,.
\end{eqnarray}
We note that in the ionized region, $T\simeq10^4\,\K$ and $\cs\simeq10\,\rm{km/s}$.
As for the direct PEW, the peaks are located at the inner edge of the outer disk.

\subsubsection{Hole size} 
\label{sec:Rhole}

Once the size of an inner hole exceeds $R\sub{crit}$, then we switch $\SigdotPE$ from the indirect (primordial) one to the direct one.

We define the hole sizes ($R\sub{hole,X}$ for X-rays and $R\sub{hole,EUV}$ for EUV) as the radius where the optical depth along the midplane is unity \citep[see, e.g., ][]{Alexander+06b, Kimura+16}.
The optical depth for the radiation with the wavelength $\lambda$ is given by $\tau_\lambda = \sigma_\lambda N\sub{mid}$, where $N\sub{mid}$ is the column density along the midplane: 
\begin{equation}
    N\sub{mid}(r)
    =\int_0^r n\sub{mid}dr'
    =\int_0^r (\rho\sub{mid}/(\mu \amu))dr'\,.
\end{equation}
$n\sub{mid}$ is the gas number density at the midplane.
We adopt the absorption cross sections $\sigma_\lambda$ of X-rays and EUV photons in the solar-metallicity case: $10^{-22}\,\rm{cm}^{-2}$ and $6.3\times10^{-18}\,\rm{cm}^2$, respectively \citep{Wilms+00, Osterbrock+Ferland06}.

\subsubsection{Smoothing functions} 
\label{sec:smthpe}

To avoid numerical problems, we smooth the $\dot{\Sigma}\sub{X,p}$ profile in the outer region, and  the $\dot{\Sigma}\sub{X,d}$ and 
$\dot{\Sigma}\sub{EUV,d}$ profiles in the vicinity of the $R\sub{hole,X}$ and $R\sub{hole,EUV}$.

The original $\dot{\Sigma}\sub{X,p}$ profile described in \citet{Owen+12} (see their Eq.\,B2) has a sharp cut-off at $100\,\au$.
This may result from the computational domain size of the hydrodynamic simulations in \citet{Owen+10}.
Hence, we make the cut-off more gradual:
The original profile contains a term $\exp\braa -\bra x/100 \ket^{10} \kett$, where $x=0.85(r/1\,\au)(\Mstar/\Msun)^{-1}$.
In this paper we modify the term to $\exp\braa -\bra x/100 \ket^{1} \kett$, which results in the lower $\dot{\Sigma}\sub{X,p}$ in $r\sim30$--100\,au and a $20\%$ decrease in $\Mdot\sub{X,p}$.

We also introduce a smoothing function in the $\dot{\Sigma}\sub{X,d}$ and 
$\dot{\Sigma}\sub{EUV,d}$ profiles following \citet{Takeuchi+05} and \citet{Alexander+Armitage07}.
After the mass-loss profile is switched to the direct PEW, there remains a small amount of gas inside the hole radius by definition.
Although the gas is expected to be heated up and flow out, both $\SigdotX$ and $\SigdotEUV$ are zero in the prescriptions of \citet{Owen+12} and \citet{Alexander+Armitage07}.
To avoid a numerical problem, we multiply the prescriptions by a smoothing function:
\begin{equation}
f_i(r)=\left[ 1+\exp\left( - \frac{r-R_{\mathrm{hole},i}}{h_{\mathrm{hole},i}}  \right) \right]^{-1}\,,
\end{equation}
where $i$ is X-ray or EUV, and $h\sub{hole}$ is the scale height at $R\sub{hole}$.

\subsection{Initial condition}\label{sec:init}

{We adopt the same initial density profile as \citetalias{Suzuki+16};}
\begin{equation}\label{eq:init}
\Sigma\sub{ini} = \Sigma\sub{1\,\au} (r/1\,\au)^{-3/2} {\exp{(-r/r_1)}},
\end{equation}
where $\Sigma\sub{1\,\au}$ is the initial surface density at $1\,\au$ and $r_1$ the initial cut-off radius.
The power law corresponds to the minimum mass solar nebula model \citep[{MMSN;}][]{Hayashi81}, which decays exponentially beyond $r_1$.
Following \citetalias{Suzuki+16}, we choose a larger value of $\Sigma\sub{1\,\au}$ than the original MMSN model: 
We adopt $\Sigma\sub{1\,\au}=1.7\times10^4\,\mathrm{g\,cm^{-2}}$ and $r_1=30\,\au$, and thus the initial disk mass, $M\sub{d,ini}=0.118\,\Msun$.

{We note that recent observations \citep[e.g.,][]{Andrews+10} showed that the surface density profile of the inner disk would be flatter ($\propto r^{-1}$, rather than $\propto r^{-3/2}$ in the MMSN model).
}
We {also} note that the wind torque depends on the initial surface density profile (see Eq.\,\ref{eq:apz}).
Therefore if a different $\Sigma\sub{ini}$ profile is assumed, the subsequent evolution of the wind-torque strength differs.
{Although we will not change the initial condition in this paper, we discuss the uncertainties in the initial condition in Sect.\,\ref{sec:discs-input}.}

\subsection{Disk dispersal condition} \label{sec:tNIR}

We will compare our results with the observations of disk lifetime, many of which use near-infrared (NIR; $\sim1$--$8\,\micron$) dust emissions \citep[see, e.g.,][]{Haisch+01,Hernandez+07,Mamajek09,Fedele+10, Yasui+10}.
Hence, following \citet[][]{Kimura+16}, we define the time when an inner disk becomes transparent in the NIR wavelength (i.e., $\kappa\Sigma < 1$ in the entire NIR-emitting region) as an inner disk lifetime ($t\sub{NIR}$).
Given the opacity $\kappa\sim10\,\mathrm{cm^2/g}$ in NIR \citep{Miyake+Nakagawa93}, the condition reads $\Sigma < 0.1\,\mathrm{g/cm^2} \equiv \Sigma\sub{crit}$.
The NIR-emitting region is defined as the region where $\Tmid \geq 300\,\K$.

In this paper we neglect the effect of dust depletion \citep[e.g.,][]{Takeuchi+05}.
The growth of small dust grains to larger bodies effectively reduces the dust-to-gas ratio that determines the NIR opacity. In other words, the dust growth would give effectively larger $\Sigma_{\rm crit}$.
We confirmed that, with the criteria $\Sigma\sub{crit}=1\,\mathrm{g/cm^2}$, $t\sub{NIR}$ becomes shorter by $\lesssim 5\%$ and therefore is not sensitive to the assumed $\Sigma\sub{crit}$ value.

\subsection{Numerical method}\label{sec:setting}
We solve the time integration of Eq.\,(\ref{eq:sgmevl}) using the time-explicit method.
The numerical flux is measured with the central difference scheme for the viscous accretion and the upwind difference scheme for the wind torque.
We confirmed the mass conservation in all calculations.
We also confirmed the numerical convergence (i.e., no dependence of the results on the Courant number).
The temperature structure including viscous heating (Eq.\,\ref{eq:Tmid}) is iteratively solved using the bisection method.

The calculation domain ranges from 0.01 to $10^4$\,au.
At the inner and outer boundaries, we impose $\pderiv{}{r}\bra \Sigma r^{-3/2} \ket =0$, which is the zero-torque boundary condition in the case that $\arp$ is constant with radius, $\apz=0$ and $T\sub{mid}\propto r^{-1/2}$ \citep[see][\citetalias{Suzuki+16}]{Lynden-Bell+Pringle74}.
The grid size is in proportion to $\sqrt{r}$ and the number of mesh points is 2000.
We stop calculations either at 20\,Myr or when a disk completely dispersed (disk mass $M\sub{disk}<10^{-10}\,\Msun$).


\begin{table}
	\centering
	\caption{Model parameters and fiducial values.}
	\label{tab:init}
  \begin{tabular}{ll}
    \hline\hline
    Parameter & Fiducial value\\
    \hline
    Stellar mass, $M_\star$ & $1\,\Msun$ \\
	Initial disk mass, $M\sub{d,ini}$ & $0.118\,\Msun$ \\
    Initial cut-off radius, $r_{1}$ & 30\,\au \\
	Stellar bolometic luminosity, $L_\star$ & $1\,\Lsun$ \\
	Stellar X-ray luminosity, $\LX$ & $10^{30}\,{\rm{erg\,s^{-1}}}$ \\
	Stellar EUV photon flux, $\Phi_{\rm{EUV}}$ & $10^{41}\,{\rm{s^{-1}}}$ \\
    \hline
  \end{tabular}
\end{table}

In Table\,\ref{tab:init}, we summarize the parameters and their fiducial values in this paper.
We adopt the typical values of the X-ray and EUV luminosities of $1\,\Msun$ T Tauri stars: $L\sub{X}=10^{30}\,\rm{erg/s}$ \citep[][]{Flaccomio+03, Preibisch+05, Telleschi+07a} and $\pEUV=10^{41}\,\rm{s^{-1}}$ \citep[][]{Bouret+Catala98, Alexander+05}.
We note that the influence of the variations in $\LX$ and the initial conditions on the evolution has been investigated in \citet{Kimura+16} (see also Sect.\,\ref{sec:LX}).


\section{Results} \label{sec:result}

Here we show the evolutionary models of protoplanetary disks varying the settings as summarized in Table\,\ref{tab:models}.
We calculated the disk evolutions with two choices of $\arp$: $8\times10^{-5}$ representing the MRI-inactive disks, and $8\times10^{-3}$ the MRI-active disks, following \citetalias{Suzuki+16}.
The MDW mass loss is also affected by the choice: $\cwo=2\times10^{-5}$ for the former, and $1\times10^{-5}$ for the latter.

\begin{table*}
	\centering
	\caption{Model settings and results.}
	\label{tab:models}
  \begin{tabular}{cccccccc} 
    \hline\hline
    Model & \multicolumn{4}{c}{Settings} & & \multicolumn{2}{c}{Results} \\ 
    \cline{2-5} \cline{7-8}
     & MDW$^a$ & MDW & PEW$^c$ & MRI$^d$ & & $t\sub{NIR}^e$ & $M\sub{acc}/M\sub{MDW}/M\sub{PEW}$ $^f$\\
     & & torque$^b$ & &  & & [Myr] & {[\%]} \\
    \hline
    \nodw-i & no & --- & on & inactive & & {15.9} & {32/---/68} \\
    \nope-i & strong & on & off & inactive&  & 14.1 & {18/76/---} $^g$  \\
    \strtrq-i & strong & on & on & inactive & & {2.98} & {9/57/35}\\
    D-i & strong & off & on & inactive & & {7.83} & {1/44/55}\\
	E-i & weak & on & on & inactive & & 5.11 & {35/23/42}\\
	\weak-i & weak & off & on & inactive & & {12.4} & {12/27/61} \\
	\hline
	A-a & no & --- & on & active & & {2.44} & {72/---/ 28}\\
	B-a & strong & on & off & active & & {8.40} & {20/76/---} $^g$  \\
	C-a & strong & on & on & active & & {1.56} & {12/67/21} \\
	D-a & strong & off & on & active & & {1.70} &  {3/75/22} \\
	E-a & weak & on & on & active & &  1.52 & {30/50/20} \\
	F-a & weak & off & on & active & & 1.70 & {23/56/21}\\
	\hline
	\end{tabular}
	\\
	\footnotesize{{\bf Notes.}
	$^{(a)}$ In the strong and weak MDW cases, Eqs.\,(\ref{eq:strongcwe})--(\ref{eq:strongfrad}) and Eqs.\,(\ref{eq:weakcwe})--(\ref{eq:weakfrad}) with $\erad=0.9$ are used, respectively (see Sect.\,\ref{sec:cw}).
	$^{(b)}$ See Sect.\,\ref{sec:apz}.
	$^{(c)}$ See Sect.\,\ref{sec:pe} for the PEW models.
	$^{(d)}$ In the MRI-inactive case 
	$(\arp, \cwo)=(8\times10^{-5}, 1\times10^{-5})$, whereas
	$(8\times10^{-3}, 2\times10^{-5})$ in the MRI-active case.
	$^{(e)}$ Inner disk lifetime (Sect.\,\ref{sec:tNIR}).
	$^{(f)}$ {Total masses, $M\sub{acc}, M\sub{MDW},$ and $M\sub{PEW}$, normalized by the initial disk mass, $M\sub{d,ini}=0.118\,\Msun$, when the disks disperse.} 
	{$^{(g)}$ In Model B (i.e., no PEW cases), although inner disks become optically thin to NIR at $t\sub{NIR}$, a small amount of gas exists even at 20\,Myr. Therefore $M\sub{acc}$ and $M\sub{MDW}$ at 20\,Myr are described.}
	}
\end{table*}

\subsection{MRI-inactive cases}\label{sec:inactive}
First we show the evolution of MRI-inactive disks, which are preferred by recent observations and theoretical studies {(see Sect.\,\ref{sec:intro})}.

\subsubsection{No MDW case (Model \nodw-i)}
\label{sec:nodw}

\begin{figure*}
	\includegraphics[width=2.1\columnwidth]{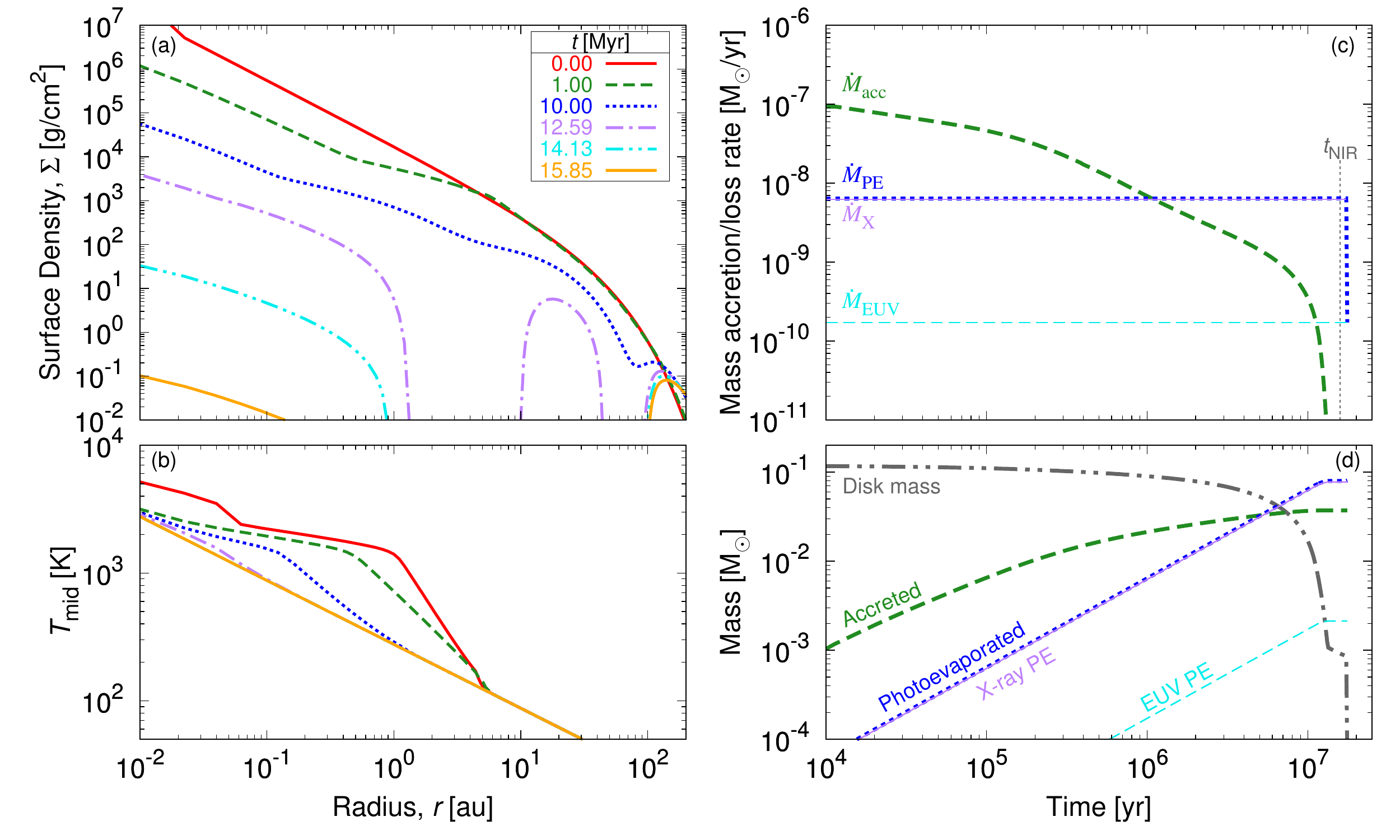}
	\caption{
    Temporal evolution of Model \nodw-i (see Table\,\ref{tab:models}).
    The panels (a) and (b) show the evolution of surface density ($\Sigma$) profile and the midplane temperature ($\Tmid$) profile, respectively.
    Each line shows a snapshot.
    The panel (c) shows the evolution of mass accretion rate ($\Mdot\sub{acc}$, thick dashed line) and mass-loss rate by photoevaporation ($\Mdot\sub{PEW}$, thick dotted), where $\Mdot\sub{PEW} = \Mdot\sub{X}+\Mdot\sub{EUV}$ (thin solid and dashed, respectively).
    We note that $\Mdot\sub{PEW} \simeq \Mdot\sub{X}$ and the two lines are overlapped.
    The thin dashed vertical line displays $t\sub{NIR}$.
    The panel (d) shows the evolution of the time-integrated masses of accretion ($M\sub{acc}$, dashed line) and photoevaporation ($M\sub{PEW}$, dotted), and the evolution of disk mass ($M\sub{disk}$, double dot-dashed line).
    The thin solid and dashed lines show $M\sub{X}$ and $M\sub{EUV}$, respectively.
    }
    \label{fig:nodw}
\end{figure*}

First let us show the results of Model \nodw-i (with the PEW but without the MDW).
Figures\,\ref{fig:nodw}a and \ref{fig:nodw}b show the evolution of the profiles of the surface density and midplane temperature.
Figure\,\ref{fig:nodw}c shows the evolution of the mass accretion rate ($\Mdot\sub{acc}$)\footnote{
In the cases with the MDW, $\Mdot\sub{acc}$ depends on the radius (see \citetalias{Suzuki+16}). In this paper we measure it at 0.01\,au.
} and the mass-loss rate by the PEW ($\MdotPE=\Mdot\sub{X}+\Mdot\sub{EUV}$).
Finally Fig.\,\ref{fig:nodw}d shows the evolutions of the disk mass, $M\sub{disk}$, the time-integrated mass of accreted materials onto the star ($M\sub{acc}\equiv\int_0^t\Mdot\sub{acc}dt'$), and the time-integrated ejected mass by the PEW ($M\sub{X}\equiv\int_0^t\Mdot\sub{X}dt'$, 
$M\sub{EUV}\equiv\int_0^t\Mdot\sub{EUV}dt'$, and
$M\sub{PEW}=M\sub{X}+M\sub{EUV}$).

The qualitative behavior of the evolution is the same as the long-term disk evolution models in the literature including both viscous accretion and the PEW  \citep[e.g.,][]{Clarke+01, Alexander+06b, Gorti+09, Owen+10, Morishima12, Bae+13, Kimura+16}:
(i) the disk mass decreases with time due to viscous accretion, (ii) a gap is created when and where the accretion rate decreases down to the PEW rate, (iii) an inner disk depletes in the viscous timescale at the gap, and then (iv) after the dispersal of the inner disk, the outer disk is directly irradiated and also quickly dispersed.

The viscous heating dominates, in particular in the inner region and in the early phase (Fig.\,\ref{fig:nodw}b).
Owing to the non-linear profile of the opacity on temperature (Eq.\,\ref{eq:kap}), the $\Tmid$ profile is also a non-smooth function of $r$ in the early phase ($\lesssim 13\,\Myr$), which results in the non-linear $\Sigma$ profile (Fig.\,\ref{fig:nodw}a).
We note that the flat $\Tmid$ profile ranging $\sim1500$--$2500\,\K$ appears because we adopted the smoothed opacity profile (Eq.\,\ref{eq:kap-smth}).

The mass-loss rate of the EUV PEW is 1--2 orders of magnitude lower than that of the X-ray PEW (see discussions in Sect.\,\ref{sec:discussion_pe}) and therefore $\Mdot\sub{PEW}\simeq\Mdot\sub{X}$.

In Model \nodw-i, the disk lifetime, $t\sub{NIR}$, is 16.1\,Myr.
Therefore the classical picture with the viscous accretion and the PEW is inconsistent with the observation in the MRI-inactive disks. 
This is because the phase (i) lasts long with the low turbulent viscosity $\arp$.
The duration of the phase (iii) is also several Myr (i.e., the inner disk lasts long), even though the inner disk is small ($\simeq3$\,au).
The amount of the total photoevaporated mass, $M\sub{PEW}$, is a factor of $\sim2$ larger than the total accreted mass, $M\sub{acc}$.

The existence of the inner disk prevents the EUV PEW from switching to the direct one. 
The X-ray PEW switches to the direct one just before the disk dispersal and then $\Mdot\sub{X}$ drops down to zero after $R\sub{hole,X}$ reaches the computational outer boundary ($10^4$\,au).

\subsubsection{No PEW case (Model \nope-i)}
\label{sec:nope}

As for the evolution of Model \nope-i (i.e., with the strong MDW but without the PEW), we refer to \citetalias{Suzuki+16} (see their Sect.\,3.2 and Fig.\,5--9).

Owing to both the MDW mass-loss and the accretion driven by the wind torque, the surface density in the inner region is significantly lower than that in Model A-i and has a positive gradient with radius.
Therefore the evolution is qualitatively different.
Due to the reduced surface density in the inner region, $t\sub{NIR}$ is smaller than that of Model A-i.

We note that even at 20\,Myr, there still remains the disk gas ({$M\sub{disk}=0.007\,{\Msun}, \Mdotacc = 2.2\times10^{-10}\,\Msun/\yr$}).
This is because the MDW is self-regulated; since the energy to launch the MDW comes from the liberated gravitational energy of accreting materials, $\MdotDW$ decreases with time along with the decreasing $\Mdotacc$.
This is a clear difference between the MDW and the PEW: the PEW does not depend on the accretion.
Therefore, only with the MDW, the disk does not disperse rapidly.
In this sense, the evolution in Model B-i is inconsistent with observations, which have suggested that $t\sub{NIR}$ is several Myr and that the entire disk disperses rapidly \citep[e.g.,][]{Andrews+Williams05}.

Nevertheless, the mass-loss rate in the early phase in Model B-i by the MDW is much larger than that by the PEW in Model A-i. Moreover in Model B-i, the total mass lost by the MDW until 20\,Myr is 4.3 times larger than that by accretion.
These express the importance of the MDW.

\subsubsection{Case with both the MDW and the PEW (Model C-i--F-i)}
\label{sec:str}

\begin{figure*}
	\includegraphics[width=2.1\columnwidth]{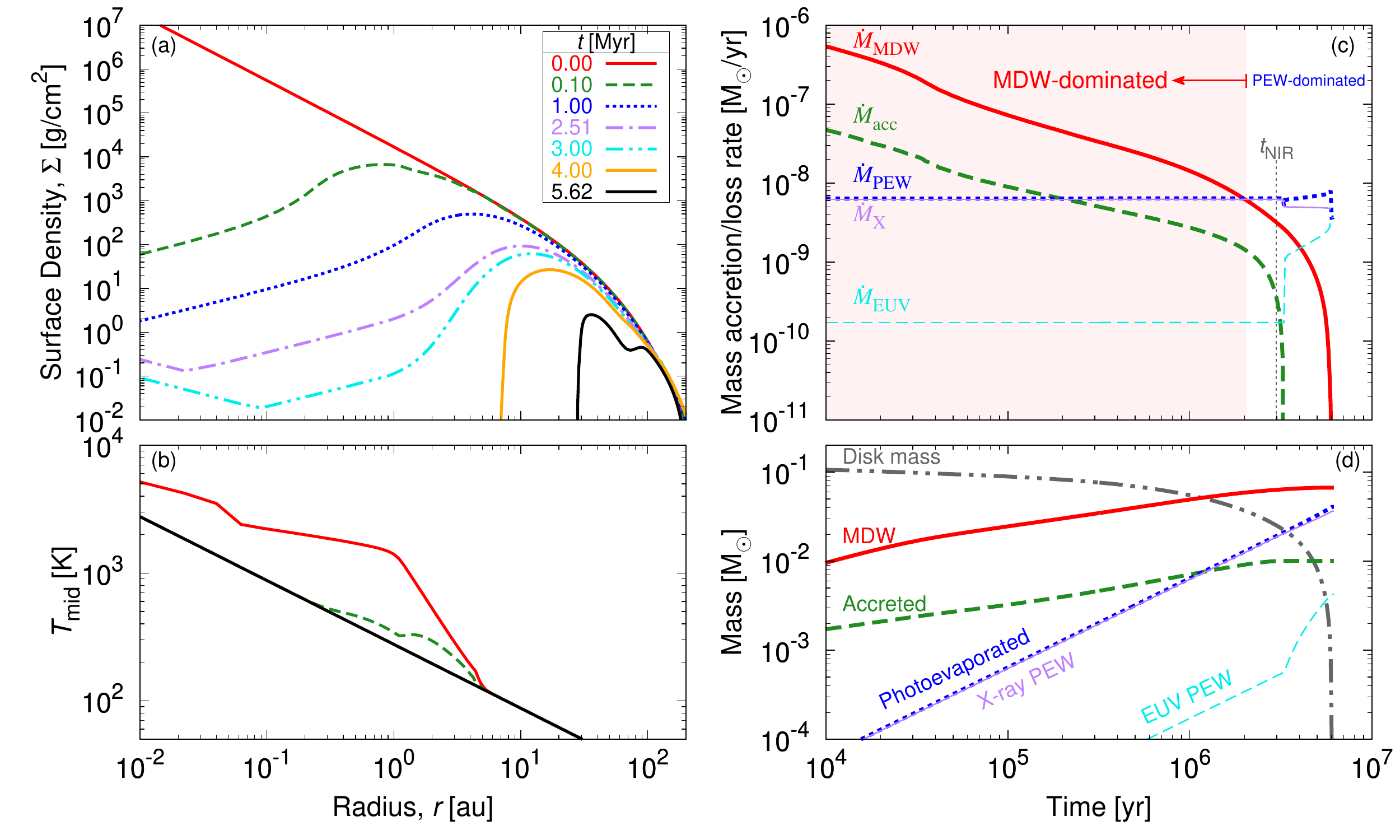}
	\caption{
    Same as Fig.\,\ref{fig:nodw} but with MDW (Model \strtrq-i).
    In the panels (c) and (d), the mass-loss rate and total ejected mass by the MDW are shown by the solid lines.
    }
    \label{fig:strtrq}
\end{figure*}

Figure\,\ref{fig:strtrq} shows the results of Model \strtrq-i, including the strong MDW and the wind torque in addition to the PEW.
The evolutionary nature is clearly different from Fig.\,\ref{fig:nodw}.

The MDW is the dominant mass-loss process especially in the early phase.
Figure\,\ref{fig:strtrq}c shows that at first $\MdotDW$ is two orders of magnitude larger than $\Mdot\sub{PEW}$, whereas $\MdotPE$ dominates after 2.0\,Myr.
In the late phase, as in Fig.\,\ref{fig:nodw}, the PEW opens a gap at 3.3\,Myr and the disk quickly disperses.
Although the PEW drives the rapid dispersal at the end, most disk materials are ejected by the MDW in Model C-i.

\begin{figure}
	\includegraphics[width=\columnwidth]{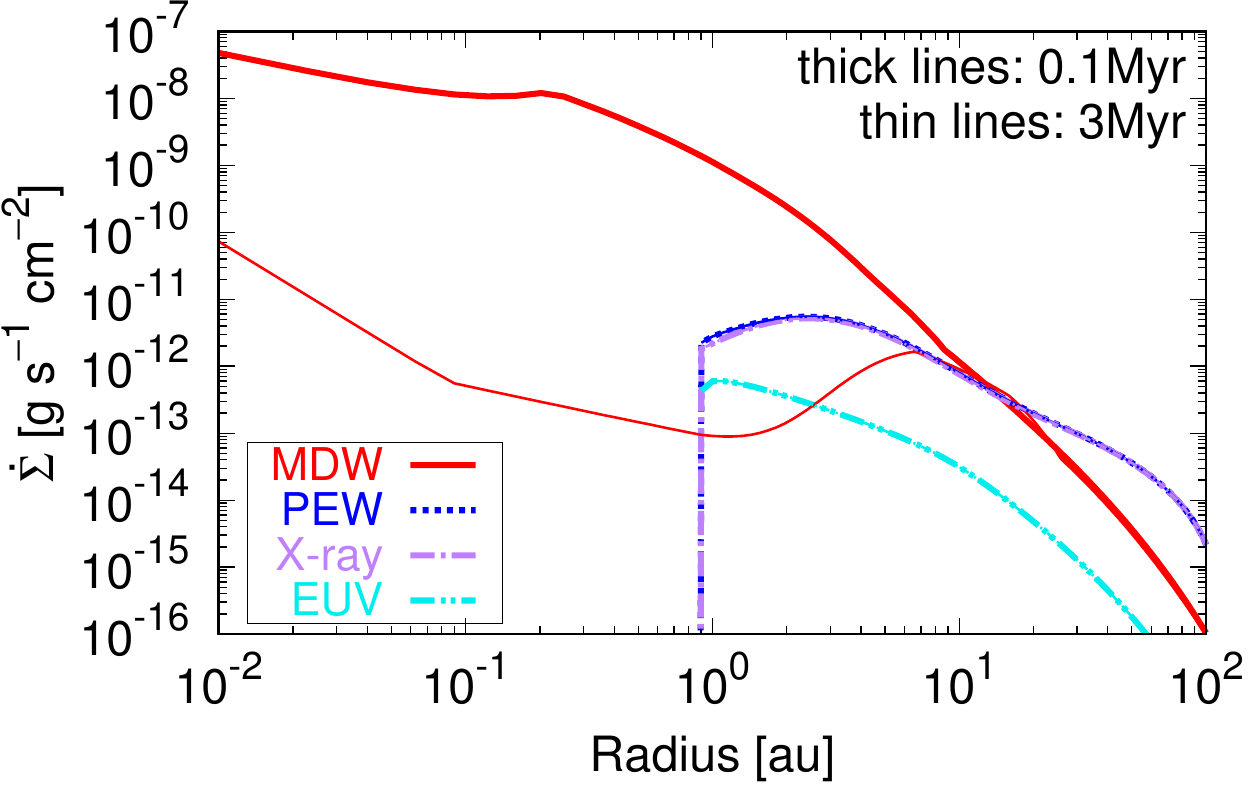}
	\caption{
    Snapshot of the profiles of $\SigdotDW$ (solid line), $\SigdotPE$ (dotted), $\SigdotX$ (dot-dashed) and $\SigdotEUV$ (double dot-dashed) at 0.1\,Myr (thick lines) and 3\,Myr (thin) in the Model \strtrq-i.
    }
    \label{fig:strtrq-Sigdot}
\end{figure}

Figure\,\ref{fig:strtrq-Sigdot} shows the mass-loss profiles of the MDW ($\SigdotDW$) and the PEW ($\SigdotPE$) at 0.1 and 3\,Myr.
$\SigdotPE$ does not change until 3.3\,Myr because it is determined by the external process of the irradiation from the central star. In contrast, the MDW dominates at the early phase and in particular in the inner region. The reason why $\SigdotDW$ dominates at the small $r$ is that the mass-loss timescale of the MDW ($\tau\sub{MDW} \equiv \Sigma/\SigdotDW$) is proportional to the Kepler time;
\begin{align}
    \tau\sub{MDW} \propto \cw^{-1}  \Omega\sub{K}^{-1}
\end{align}
(see Eqs.\,\ref{eq:h}--\ref{eq:DW}; \citetalias{Suzuki+16}).

We note that interestingly $\tnir$ (3.0\,Myr) is shorter than the gap-opening time.
The wind torque maintains the accretion rate at a high level ($\sim10^{-10}\,\Msun/\yr$ at the inner edge) even with $\Sigma<0.1\,\mathrm{g/cm^2}$ in the inner disk.
This vigorous accretion (i.e., large radial velocity) feeds materials into a few au and prevents the PEW from opening a gap even when the inner disk is transparent to NIR.

In Model\,\str-i, the wind torque is not considered. In this case the disk lifetime $\tnir$ ($=7.9\,\Myr$) is much longer than that of Model \strtrq-i (3.0\,Myr).
This difference clearly illustrates the great impact of the wind torque in the MRI-inactive case.

The cases with the weak MDW give a quantitatively longer $\tnir$ and a smaller $M\sub{MDW} (\equiv \int_0^t\Mdot\sub{MDW}dt')$ than the strong MDW cases. However, the qualitative behavior of the evolution is the same.

We note that the dominance of $M\sub{MDW}$ over $M_{\rm acc}$ does not necessarily mean that a large portion of the gas is lost by the MDW rather than the accretion.
This is because it is expected that a sizable fraction of the wind material launched from the inner disk does not escape from the system but accretes onto the central star via funnel-wall accretion \citep{Takasao+18}, which we do not take into account in our model. In other words, a part of $M\sub{MDW}$ is regarded to contribute to the accretion from an observational point of view.

\begin{figure}
	\includegraphics[width=\columnwidth]{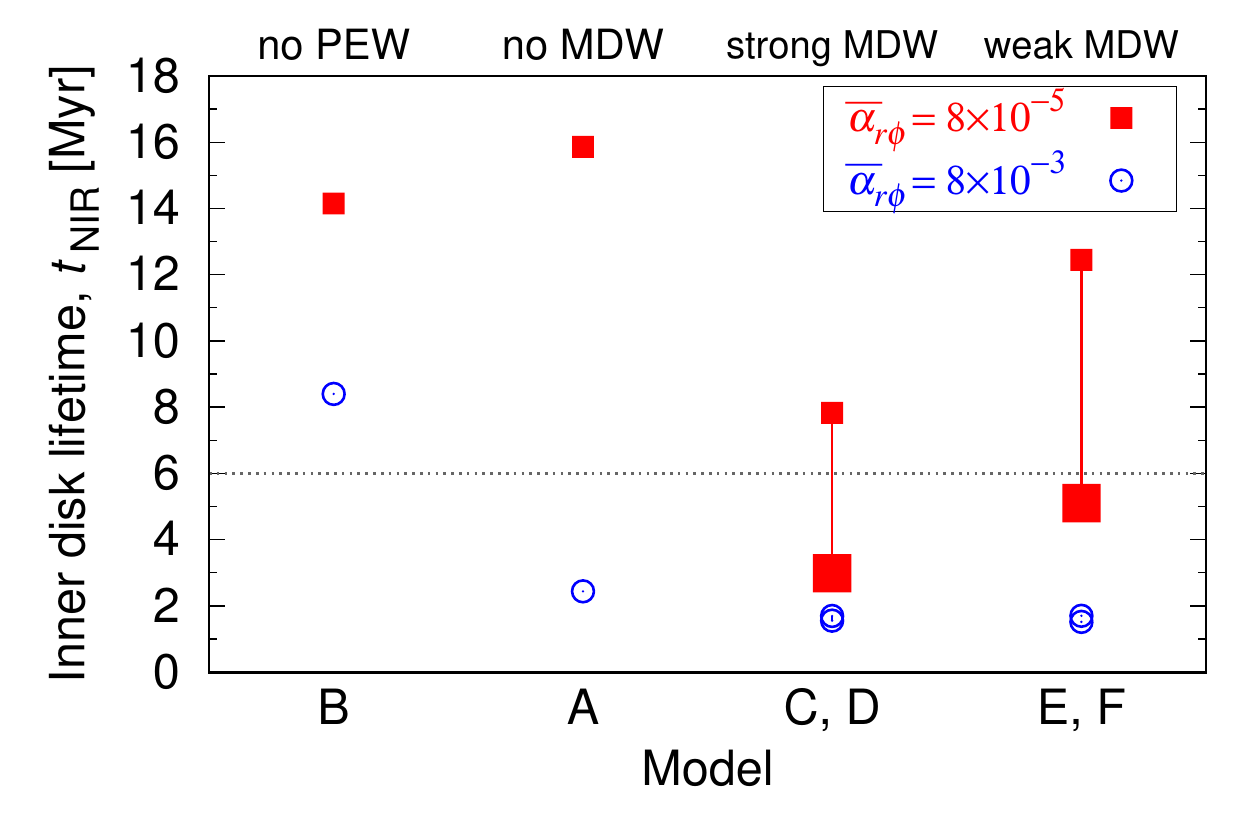}
	\caption{
    Comparison of the inner disk lifetime, $t\sub{NIR}$, in the MRI-inactive (squares) and MRI-active (circles) cases.
    The differences between Models C and D and between E and F illustrate the impact of wind torque (see also Table\,\ref{tab:models}).
    The horizontal dotted line (6\,Myr) indicates a constraint from NIR observations \citep[e.g.,][]{Haisch+01}.
    {The symbols of Model C-i and E-i are highlighted by a larger filled square because they are consistent with observations (see text).}
    }
    \label{fig:tNIR}
\end{figure}

Finally, in Fig.\,\ref{fig:tNIR} we summarize the inner disk lifetime $t\sub{NIR}$.
In the MRI-inactive cases, $t\sub{NIR}$ depends on the MDW model parameters.
Under the current settings, we find that $t\sub{NIR}$ is comparable to the observed values ($\sim$3--6\,Myr), only if the PEW, the MDW and the wind torque cooperatively operate {(i.e., Model C-i and E-i)}.

\subsection{MRI-active cases}\label{sec:active}
Here, we describe the evolution of MRI-active disks.
The qualitative behavior of the disk evolution is similar to the MRI-inactive cases. In Model C-a--F-a, most materials are dispersed by the MDW and then the PEW plays a role during the final phase.
However, the quantities are different from the MRI-inactive cases.

The MRI-active cases with the larger $\arp$ result in the more vigorous accretion than the MRI-inactive cases.
Thus $M\sub{acc}$ is larger in all the MRI-active cases (Model A-a--F-a, see Table\,\ref{tab:models}).
As a consequence, the MDW is also stronger and therefore $M\sub{MDW}$ is also larger in the MRI-active cases, because the MDW is launched by using the accretion energy.
Regardless of the MDW model, $M\sub{MDW}>M\sub{PEW}$ in MRI-active cases.

Despite the larger impact of the MDW, the disk lifetime $t\sub{NIR}$ is less sensitive to the MDW model in the MRI-active cases: it is around 2\,Myr except for Model B-a.
Without the PEW, even if a large $\arp$ is assumed and the strong MDW is adopted with the wind torque, $t\sub{NIR}$ is longer than the observed value.

\section{Discussions: Model caveats} \label{sec:discussion}

\subsection{Photoevaporation models}
\label{sec:discussion_pe}

In this work we adopt the X-ray and EUV PEW models in the literature (see Sect.\,\ref{sec:pe}).
Recently there has been much progress on the PEW models by \citet{Wang+Goodman17} and \citet{Nakatani+18a,Nakatani+18b}.
They performed radiation hydrodynamic simulations including X-ray and UV radiation.
They claimed that the X-ray PEW rate by \citet{Owen+10,Owen+11b} was overestimated, because Owen et al. did not self-consistently solve radiative transfer and thermochemistry.
Instead, they claimed that the UV (EUV and far-UV) PEW drives the mass-loss comparable to the X-ray PEW rate obtained by \citet{Owen+10} (i.e., $10^{-8}$--$10^{-9}\,\Msun/\yr$).
Moreover, the overall profile of the PEW profile (i.e., decreasing with radius) is not changed \citep[see Fig.\,8 of][$\SigdotPE\propto r^{-2}$]{Wang+Goodman17}.
Therefore we believe that the the evolutionary nature and the conclusion in this article are not affected by the updated PEW model. 
The long-term disk evolution with the recent PEW model should be studied in future.

\subsection{Interplay between photoevaporative and MHD winds}
\label{sec:interplay}

In this paper we simply assume that the total mass-loss rate is the sum of the PEW and the MDW, that is, $\dot{\Sigma} = \SigdotPE + \SigdotDW$.
However, the interplay may not be so simple.
Recently \citet{Wang+19} have performed the global radiation magnetohydrodynamic simulations for the first time.
They found that adding EUV photons reduces the mass-loss rate due to the enhanced ambipolar dissipation (see their Sect. 5.4).

As shown in Fig.\,\ref{fig:strtrq}c, $\MdotDW$ and $\MdotPE$ are comparable only for a short period of the whole lifetime.
In the rest of the time, one mechanism dominates the other; the MDW (PEW) dominates the PEW (MDW) in the early (late) phase.
Therefore we expect that the interplay may not affect the results in this article.
Further investigation on the magneto-thermal winds is highly encouraged.

\subsection{Uncertainties of input parameters}\label{sec:discs-input}

We adopt a unique mass and size for the initial conditions in all cases.
These are also known to affect the disk lifetime \citep[e.g.,][]{Alexander+Armitage09}.
Therefore, the disk lifetime in Fig.\,\ref{fig:tNIR} should be regarded as a median value, but in reality it includes a large scatter.

The turbulent viscosity, the MDW mass-loss rate, and the wind torque depend on the strength and shape of the magnetic field, which are still uncertain. Therefore the values of $\arp$, $\cw$ and $\apz$ also remains uncertain, even though they largely affect the disk evolution (see Sect.\,4.1 of \citetalias{Suzuki+16}).
To construct realistic disk evolutionary models and predict the disk lifetime, it is essential to understand the global evolution of the poloidal magnetic field in disks.

To investigate the influence of the uncertainties in $\arp$, $\cw$ and $\apz$ on the disk evolution, we perform a suit of disk evolution calculations with varying them.
We use a Monte Carlo approach to derive the $\arp$ and $\apz$ values.
We assume a Gaussian distribution with the mean values to be $\arp=8\times10^{-4}$ and $\apzo = 10^{-5}$ (see Eq.\,\ref{eq:apz}), and the standard deviation of 1\,dex.
According to \citet[][]{Suzuki+10}, $\cw$ has a weak positive correlation with  $\arp$. Therefore we derive $\cwo$ by using both the $\arp$ value and the linear fit of the two cases in this paper, that is, $(\arp, \cwo) = (8\times 10^{-5}, 1\times 10^{-5})$ and $(8\times 10^{-3}, 2\times 10^{-5})$. We set a maximum value of $\cwo$ to be $5\times10^{-5}$.

For the Monte Carlo simulations, we need to save the CPU time of each simulation. Since we use the time-explicit method, we increase the time step by enlarging the grid sizes; the number of mesh points and the calculation domain are changed to 200 from 2000, and to [0.1, $3\times10^3$]\,au from [0.01, $10^4$]\,au, respectively.
We performed the simulation of Model C-i with the coarser grids and confirmed that the change of the disk lifetime, $t\sub{NIR}$, was 5.7\%.

\begin{figure}
	\includegraphics[width=\columnwidth]{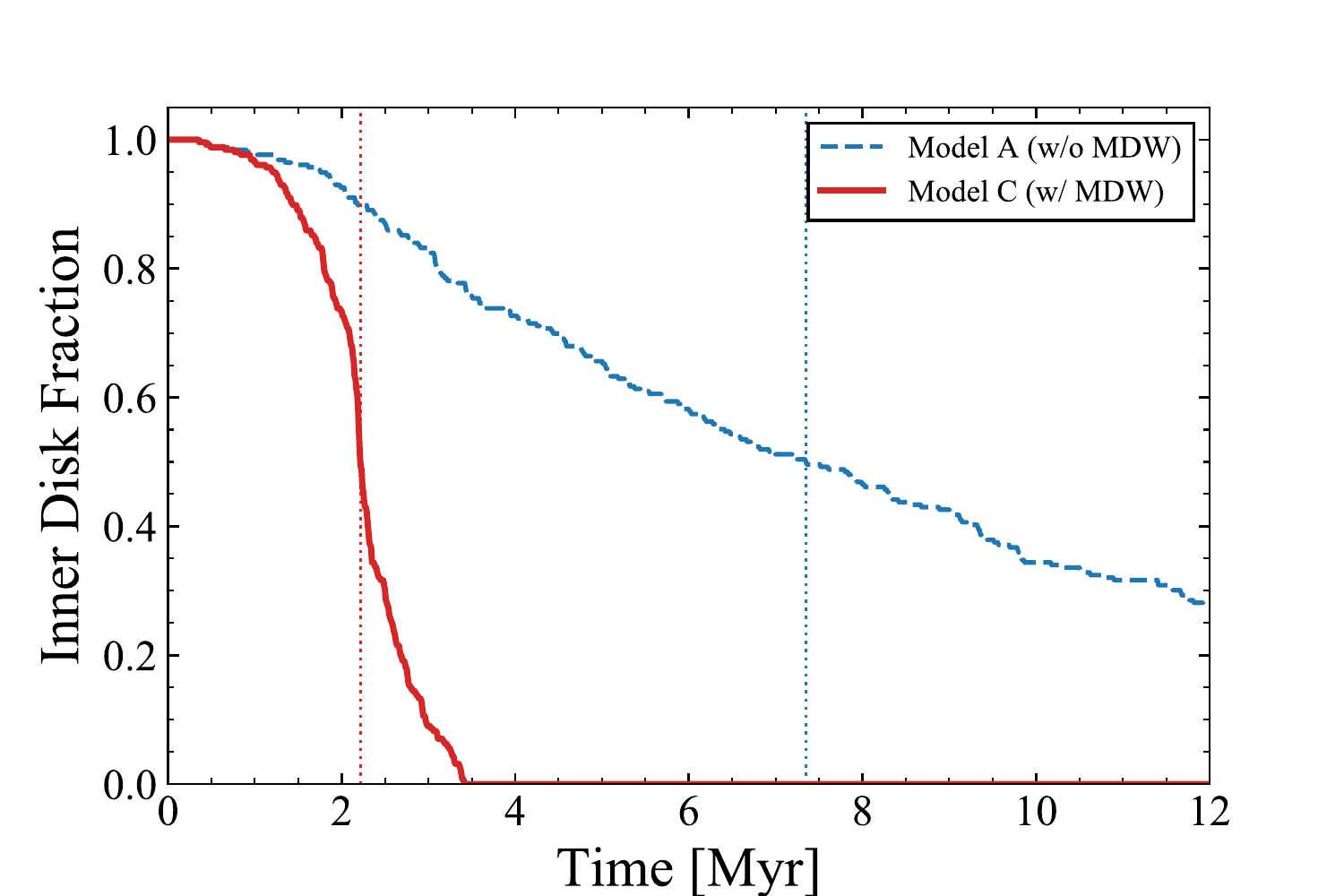}
	\caption{
    Inner disk fraction as a function of time.
    The solid and dashed lines show the results of two Monte Carlo simulations in Model C (i.e., with both MDW and PEW) and Model A (i.e., without MDW but with PEW), respectively.
    The vertical dotted lines illustrate their half-life period (2.22 and 7.35\,Myr, respectively).
    }
    \label{fig:MC}
\end{figure}

Here we focus on the two cases: Model A (i.e., with PEW but without MDW; see Model A-i and A-a in Table\,\ref{tab:models}) and Model C (i.e., with both PEW and MDW). We performed 256 calculations for each case.
Figure\,\ref{fig:MC} shows the inner disk fraction (i.e., the fraction of disks with $t\sub{NIR}\geq t$) as a function of time, $t$.
We find that the variations in $\arp$ and $\apz$ result in the gradual decrease of the disk fraction in both cases.
We find a clear difference; the half-life period is 2.22\,Myr in Model C and 7.35\,Myr in Model A.
The observed half-life period is a few Myr \citep[see][]{Haisch+01, Mamajek09} and therefore seems to prefer Model C.

We note that, however, we find that the observed inner disk fraction \citep[see][]{Mamajek09, Yasui+14} is broader than that of Model C. This may point out the importance of the variety in the other parameters, such as the initial mass distribution and $\LX$.
We will investigate this issue in our future work.

\subsection{Effect of stellar evolution}\label{sec:LX}

\citet{Kimura+16} reported that the X-ray luminosity, $\LX$, has an even larger impact on the disk lifetime than the initial conditions.
Although so far we have adopted a constant $\LX=10^{30}\,$erg/s, the young star's $\LX$ is known to evolve with time \citep[see, e.g.,][]{Flaccomio+03}.
Here we investigate the impact of the $\LX$ evolution on the disk evolution, with the same parameters as Model C-i except for $\LX$.

The young stars' $\LX$ is well correlated to the stellar bolometric luminosity $\Lstar$ and therefore we can safely approximate that the ratio $\LX/\Lstar$ is constant \citep{Noyes+84,Mangeney+Praderie84}.
Following \citet{Wright+11}, we adopt $\LX/\Lstar=10^{-3.13}$.
Also, we simulate {the evolution of} a $1\,\Msun$ pre-main-sequence star using the MESA code \citep{Paxton+11} \citep[version 2258, see the details in][]{Kunitomo+11}.
We assume that the star enters its pre-main sequence (i.e., we set $t=0$) on the birthline introduced in \citet{Stahler+Palla05}.
Together with the $\LX/\Lstar$ ratio and the $\LX$ evolution model, we obtain the $\LX$ evolution.

\begin{figure}
	\includegraphics[width=\columnwidth]{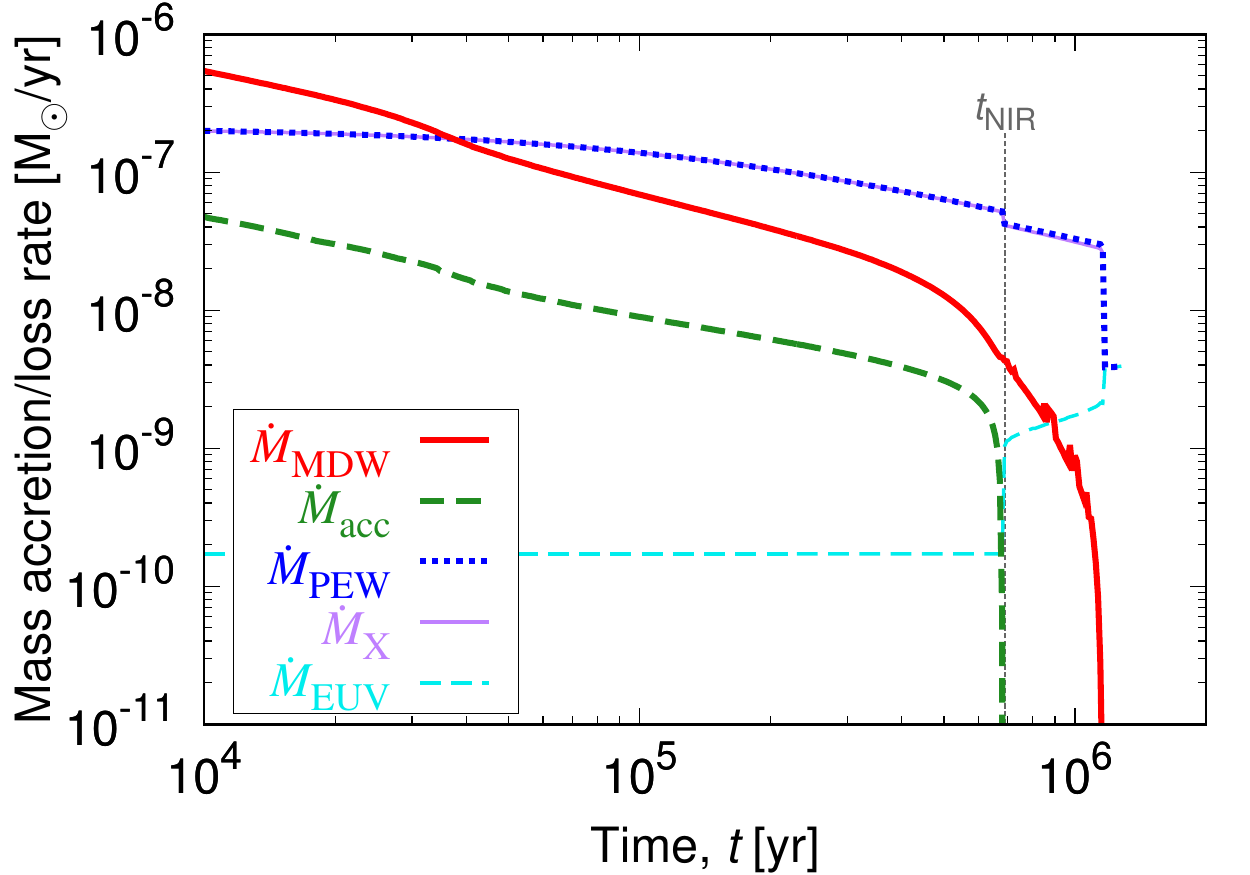}
	\caption{
    Same as Fig\,\ref{fig:strtrq}c, but with time-dependent $\LX$.
    }
    \label{fig:LX}
\end{figure}

Figure\,\ref{fig:LX} shows the disk evolution with the time-dependent $\LX$.
$\MdotPE$ gradually decreases with time, in contrast to the original Model C-i case which gives a constant $\MdotPE$.
However, its time evolution is much slower than that of $\Mdotacc$ and $\MdotDW$. 
Therefore we conclude that our results above (i.e., the MDW dominates in the early phase and the PEW in the late phase) are valid.
Since the evolution of lower-mass stars is slower, the effect of the $\LX$ evolution can be safely neglected for the disk evolutions around low-mass ($\lesssim 1\,\Msun$) stars.

\section{Conclusions}
\label{sec:conclusion}

We investigated the long-term disk evolution considering the viscous accretion, the PEW mass loss, and the mass loss and wind torque by the MDW.
We performed simulations varying the wind models ($\SigdotPE$ and $\SigdotDW$), the strength of viscosity ($\arp$) and the effect of the wind-driven accretion ($\apz$).
In particular, we focused on MRI-inactive (i.e., low $\arp$) disks suggested by recent observations and theoretical studies {(see Sect.\,\ref{sec:intro})}.
We started the calculations from the early phase of disk evolutions and thus assumed a relatively compact (cut-off radius $r_1=30\,\au$), massive ($0.118\,\Msun$) disk around a $1\,\Msun$ star.

We found that the MDW (PEW) dominates the PEW (MDW) in the inner (outer) disk in the early (late) phase.
Each wind process has a distinct role on the disk evolution.
In particular, in the MRI-inactive cases, both the MDW with the wind torque and the PEW mass loss are required to work in a cooperative manner to explain the observed inner disk lifetime (i.e., $t\sub{NIR}\lesssim$ several Myr).
{We confirmed the necessity of the cooperation of MDW and PEW using Monte Carlo simulation with varying $\arp$, $\cw$ and $\apz$.}
In the MRI-active cases, although $t\sub{NIR}$ is insensitive to the adopted MDW model, 
the MDW eject most materials (Table\,\ref{tab:models}) and 
affect the density profile.

The effect of stellar evolution on the disk evolution has not been investigated in the previous works.
We confirmed that, at least around low-mass ($\lesssim1\,\Msun$) stars, it can be safely neglected.

{In this work we have regarded protoplanetary disks as a one-component fluid, that is, we have not considered dust grains or the abundances of each element. 
It is important to model the evolution of disk composition not only for the evolution itself \citep[e.g.,][]{Gorti+15} but also the properties of formed planets \citep{Guillot+Hueso06}.
}

%
In this work we have not surveyed large parameter ranges of the initial conditions, $r_1$ and $M\sub{d,ini}$, the stellar X-ray luminosity, $\LX$, and the stellar mass, $\Mstar$. 
There still remain unresolved problems concerning the evolution of protoplanetary disks and transitional disks; (i) the decreasing disk fraction as a function of the cluster age \citep[e.g.,][]{Haisch+01,Mamajek09}, (ii) the dependence of the disk lifetime on stellar mass \citep[e.g.,][]{Hillenbrand+92,Yasui+14}, and (iii) the fraction, accretion rate and hole size of transition disks \citep[e.g.,][]{Owen16}.
Simulations in a wide parameter space are required to directly compare our model results to these observational constraints \citep[e.g.,][]{Alexander+Armitage09, Kimura+16}, which we will pursue in our future works. 
We are planning to investigate the dependence of the disk evolution on the initial conditions by incorporating a disk formation model \citep[e.g.,][]{Takahashi+13}.

\section*{Acknowledgements}

We are grateful to Hiroshi Kobayashi, Shinsuke Takasao, Shoji Mori, Kei E. I. Tanaka, and Ryunosuke Nakano for fruitful discussions and comments.
{We appreciate the constructive comments of the anonymous referee, which helped us to improve this paper.}
This work was supported by JSPS KAKENHI Grant Numbers 23244027, 16H02160, 17H01105 and 17H01153.
{This work made use of the IPython package \citep{Perez+Granger07}, matplotlib, a Python library for publication
quality graphics \citep{Hunter07}, and NumPy \citep{vanderWalt+11}.}

{\textit{Software}: Numpy}




\bibliographystyle{mnras}




\appendix



\bsp	
\label{lastpage}
\end{document}